\documentclass[aps,showpacs,prl,twocolumn,superscriptaddress]{revtex4}
\usepackage{graphicx}
\usepackage{amssymb}
\usepackage{amsmath}
\usepackage{psfrag}

\newcommand{\dert}{\partial_t}
\newcommand{\derx}{\partial_x}

\begin{document}

\title{Quantum decay  of dark solitons}
 \author{D.M.~Gangardt}
\affiliation
{School of Physics and Astronomy, University of Birmingham, Edgbaston,
Birmingham, B15 2TT, UK  \email[e-mail: ]{d.m.gangardt@bham.ac.uk}
}
\author{A. Kamenev }
 \affiliation{Fine Theoretical Physics Institute and 
Department of Physics, University of Minnesota,
   Minneapolis, MN 55455 }



\begin{abstract}
Unless protected by the exact integrability, solitons are subject to
dissipative forces, originating from a thermally fluctuating background.
At low enough temperatures $T$ background fluctuations should
be considered as being quantized which enables us to calculate finite lifetime
of the solitons $\tau\sim T^{-4}$. We also find that the coherent nature of
the  quantum fluctuations
leads to long-range interactions between the solitons mediated by
the superradiation. Our results are of relevance to current experiments
with ultracold atoms, while the approach may be extended to solitons in
other media.
\end{abstract}

\maketitle

Soliton dynamics is in a heart of multiple areas of physics and applied
mathematics \cite{Konotop2004Landau}.
The recent resurgence of interest in  propagation of solitons through a
dynamic media \cite{Muryshev2002Dynamics,Jackson2007}
was stimulated by a number of experimental observations
\cite{Burger1999Dark,Becker08} of dark solitons (DS) in  one
dimensional (1D) Bose systems. The studies of the soliton dissipative dynamics
\cite{Dimensional_crossover,Muryshev2002Dynamics,Jackson2007}
are probably the most relevant for the current experiments. Indeed the record
lifetime of the dark soliton is $\sim 200$ms \cite{Becker08},
calling for understanding of the ultimate quantum limits for soliton
persistence.

As follows from the Mermin-Wagner theorem, the mean-field
approach is not valid in 1D even at zero temperature. This calls for an
essentially quantum description, first achieved by Lieb and Liniger
\cite{Lieb_Liniger_1963} for 1D bosons with the contact two-body
interactions, 
solvable  within the Bethe ansatz.  As
was latter shown by Kulish, Manakov and Faddeev \cite{KulishManakovFaddeev76}
one of the fundamental excitations within the Bethe ansatz classification (so
called Lieb II mode \cite{Lieb_1963}) essentially coincides with DS. This fact
preserves existence of DS within the Lieb-Liniger model even on the quantum
level. The underlying reason for such a stability is the infinite set of
conservation laws, characteristic for exactly integrable models
\cite{FaddeevTakhtajanBook}. 
It allows exact canonical transformation to  collective coordinates 
of the soliton and its description as a point-like particle obeying 
quantum mechanics \cite{Dziarmaga}. At $T=0$
the Heisenberg uncertainty of the soliton position and its quantum 
spreading during the evolution impedes the imaging of dark solitons 
\cite{Moelmer}.

At finite $T$ the soliton loses its quantum coherence and
it was realized in Ref.~\cite{Muryshev2002Dynamics} that the major factor
limiting DS lifetime is the lack of the exact integrability.  In a strongly
confined (with transverse frequency $\omega_\perp$) gas it stems from the
virtual transitions to higher states of the transverse quantization leading to
the effective three-body contact interactions with the coupling constant
$\alpha = -12\ln(4/3)\, g^2 /(\hbar\omega_\perp)$ \cite{Mazets2008Breakdown}.
Here $g$ is the strength of delta-like two-body interaction.  On the classical
level such three-body interactions lead to a cubic in density term $\delta L =
-\alpha n^3/6$ in GP Lagrangian \cite{Muryshev2002Dynamics}.  The
corresponding modified Gross-Pitaevskii (GP)
equation still possesses a one parameter family of
soliton solutions, however, due to the broken classical integrability, such
solitons are unstable against scattering on thermally excited density
fluctuations.  This classical mechanism of soliton dissipation was shown to
result in a finite lifetime $\tau\sim (T/\mu)^{-1}$, for temperatures $T$
higher than chemical potential $\mu$ of 1D liquid~\cite{Muryshev2002Dynamics}.
\begin{figure}[t]
  \includegraphics[width=6.5cm]{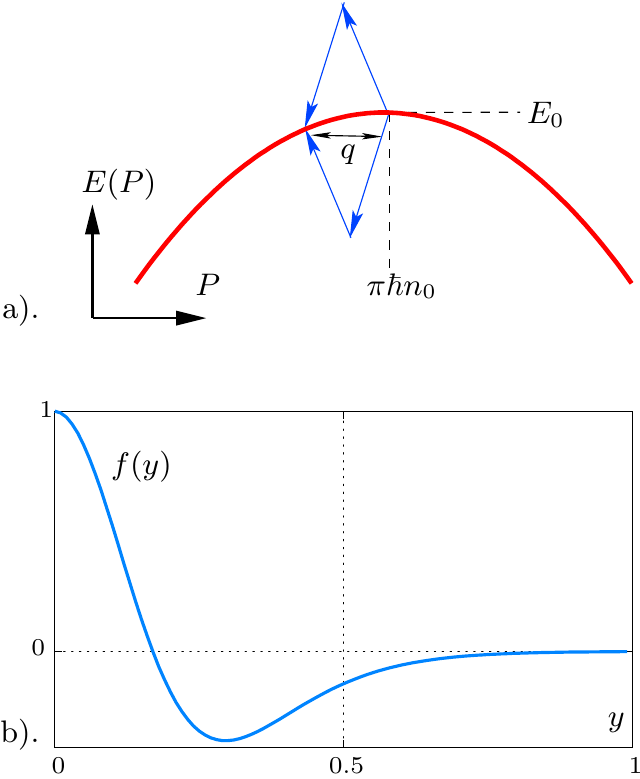}
  \caption{a) Dispersion relation $E(P)$ of soliton
    and   two-phonon processes  leading to the dissipation.
    The arrows represent absorption and emission of
    long wavelength phonons; their slope is given by sound velocity $c$.
    b) Function $f(y)$ defined in Eq.~(\ref{eq:f}).  }
  \label{fig:dispersion}
\end{figure}
In this Letter we report a study of the soliton dynamics in the
low-temperature regime $T\ll \mu$.  
For such low temperatures the phonons are
``cold'': the  density and phase
fluctuations  of the background need to be treated as quantized objects. 
Since the soliton velocity $V$ is always
smaller than the speed of sound $c$, 
emission of a single phonon is forbidden by the energy and
momentum conservation, \emph{i.e.} by Landau criterion. The leading allowed
process is the Raman two-phonon scattering depicted in
Fig.~\ref{fig:dispersion}a, where one thermal phonon is absorbed and another
one reemitted.
The rate of these processes 
and their physical consequences in the dynamics of DS
constitutes our main results which are
summarized in Eqs.~(\ref{eq:vdot}-\ref{eq:tau}) followed by their derivation.

For small soliton velocities $V\ll c$ the characteristic momentum
transferred to phonons  $q\sim T/c$ is much less than the typical DS momentum
$\pi\hbar n_0$ justifying the semilassical treatment and we find the equation
of motion 
\begin{eqnarray}
  \label{eq:vdot} M^* \dot V = -\kappa V\,.
\end{eqnarray} Here $M^*=-4\hbar n_0/c$ is the (negative) effective mass of
the soliton, while the right hand side is the viscous friction force due to
phonon scattering \cite{foot1}.  Due to the negative effective mass $M^*$
the peculiarity of DS dynamics is that the friction
force accelerates rather than decelerates the soliton. The accelerated
DS looses its energy and eventually thermalizes with the phonons.
The quantum nature of the  phonons
manifests itself in long-range interactions between solitons if one
considers a generalization of Eq.~(\ref{eq:vdot}) for a gas of
DS with the set of coordinates $X_i$
and velocities $V_i$,
\begin{equation}\label{eq:soliton-gas} M^*\dot V_i = -\kappa(T)
\sum_j\, \frac{V_i+V_j}{2}\, f\left(\frac{X_i-X_j}{L_T}\right)\,,
\end{equation} where the temperature length is defined as $L_T=\hbar c/ T$ and
the function $f(y)$ is defined in Eq.~(\ref{eq:f}) and shown in
Fig.~\ref{fig:dispersion}b.  It is characterized by $f(0) =1$ and $f(y)\approx
-30\pi y e^{-4\pi y}$ for $y\gg 1$. As a result, only solitons within the
distance $X_i-X_j\lesssim L_T$ interact effectively with each other.  This is
clearly only possible deep in the quantum regime $T\ll \mu $, where $L_T$ is
much larger than DS size given by the healing length $\xi=\hbar/mc$.
Contrary to the usual potential forces, the mutual
friction in Eq. (\ref{eq:soliton-gas}) is an {\em even} function of the
relative coordinate, and an {\em odd} function of the center of mass velocity.
As a result it does not affect the relative motion $M^*(\dot V_1-\dot V_2)
=\kappa (V_2-V_1)$ while the center of mass motion is strongly affected
$M^*(\dot V_1+\dot V_2) = -\kappa (V_1+V_2)[1+f((X_1-X_2)/L_T)]$.  In case
$X_1-X_2\lesssim 0.2 L_T$ the two DS accelerate almost twice as fast as a
single one.  This is a consequence of the superradiation, when phonons emitted
by one DS stimulate coherent emission by the other.
The same mechanism slows down the acceleration of two DS center
of mass for  $X_1-X_2\gtrsim  0.2 L_T$, Fig.~\ref{fig:dispersion}b.

We found for the viscosity coefficient
\begin{eqnarray}
  \label{eq:visc_low} \kappa(T) = \frac{1024 \pi^3}{1215} \frac{\alpha^2
n_0^4}{\hbar\, c^2} \left(\frac{T}{\mu}\right)^4\,;\quad\quad T\ll \mu\,.
\end{eqnarray}
Its sublinear dependence on temperature reflects new regime of ``cold''
degenerate phonons.  In the opposite limit $T\gg \mu$ we find linear law
$\kappa \sim (\alpha^2n_0^4/\hbar c^2)(T/\mu)$ in agreement with
Ref.~[\onlinecite{Muryshev2002Dynamics}], which is a result of the soliton
interacting with ``hot'' classical phonons.
The typical lifetime \cite{footvc} of DS may be estimated from
Eq.~(\ref{eq:vdot}) as
\begin{eqnarray}
  \label{eq:tau}
 \tau =
\frac{|M^*|}{\kappa} = \frac{1215}{256\pi^3}
\frac {\hbar^2 c}{\alpha^2 n_0^3}
\left(\frac{\mu}{T}\right)^4 .
\end{eqnarray}
Taking $^{87}$Rb atoms confined with transverse frequency
$\omega_\perp = 1$kHz we find  $g = 10^{-39}\;\mathrm{J\  m}$,
and  $-\alpha = 12 \log (4/3)\,
g^2/\hbar\omega_\perp \simeq 3.5\cdot 10^{-47}\; \mathrm{J\ m^2}$,
which for  density $n_0 =  10^8 \;\mathrm{m}^{-1}$ and sound velocity
$c\simeq 1  \;\mathrm{mm}/\mathrm{s}$
yields  $\tau \simeq 100$ms$\times(\mu/T)^4$. Experimentally $\mu/T \sim 1$
so our results are in
a reasonable agreement  with the observed \cite{Becker08} soliton life-time of
$\tau \sim 200$~ms.

We first discuss the nature of the soliton-phonon interactions. To this end we
use the integrable case without the cubic in density
$n^3$ term in the Lagrangian, and verify
the absence of phonon emission by DS.  We then reintroduce $\alpha$
and derive the dissipative force
Eqs.~(\ref{eq:vdot}), (\ref{eq:visc_low}).  The weakly interacting Bose gas
may be described by the Lagrangian
\begin{equation}
  \label{eq:action}
  L \!= \!\int\!dx \Big[
    \phi \,\dert n
    - \frac{\left(\derx \sqrt{n}\right)^2}{2m}\, -\frac{n \left(\derx
      \phi\right)^2}{2m}\, -
  \frac{g}{2} (n-n_0)^2  \Big],
\end{equation}
where $n(x,t)$, and $\phi(x,t)$ are  density and phase fields.
The gradient of the  latter is related to the superfluid velocity $u(x,t) =
\derx \phi(x,t)/m$, where $m$ is the mass of the bosons.
In Eq.~(\ref{eq:action}) we have subtracted the contribution of
uniform density profile $n_0$.
Variation of the Lagrangian (\ref{eq:action}) with respect to the fields gives
the equations of motion
\begin{eqnarray}
  \label{eq:eqs_of_motion_n}
  \dert n  &=& - \derx \frac{n}{m}\, \derx \phi\,; \\
  \label{eq:eqs_of_motion_phi}
  -\dert \phi &=& -\frac{\derx^2
  \sqrt{n}}{2m \sqrt{n} }
   + \frac{\left(\derx \phi\right)^2}{2m}  + g(n-n_0)\,.
\end{eqnarray}
This is the hydrodynamic form of GP equation \cite{PitaevskiiStringariBook},
Eq.~(\ref{eq:eqs_of_motion_n}) is the continuity equation, while
Eq.~(\ref{eq:eqs_of_motion_phi}) is the Euler equation.
In the long wavelength limit, they reduce to the wave equation describing low
energy sound waves (phonons) propagating with the sound velocity $c$.
The latter  depends on the background density $n_0$ and can be
obtained from  the thermodynamic  relation
$mc^2/n_0 =\partial \mu/\partial n_0=g$.
The corresponding Lagrangian written in terms of small
variations of the density $\rho(x,t)=n(x,t)-n_0$ becomes
\begin{eqnarray}
  \label{eq:action_luttinger}
  L_\mathrm{ph} = \int \!  dx \left[
    \phi\dert \rho - \frac{n_0+\rho}{2m}\,  (\derx \phi)^2
  - \frac{mc^2}{2n_0}\, \rho^2\right].
\end{eqnarray}
The nonlinear phonon interactions term $\rho(\derx \phi)^2$
is retained  since it is going to play an important role below.

In addition to the phonon modes, governed by Eq.~(\ref{eq:action_luttinger}),
equations of motion~(\ref{eq:eqs_of_motion_n}),~(\ref{eq:eqs_of_motion_phi})
support a one parameter family of soliton solutions
\cite{PitaevskiiStringariBook}  with the localized
profiles of density $n_\mathrm{s} (x-Vt;V)$ and current $u_\mathrm{s}(x-Vt;V)$
given by
\begin{eqnarray}
n_\mathrm{s}  (z;V) &=&
n_0\left(1- \frac{\sin^2 (\Phi/2)}
{\cosh^2 \big(z mc \sin(\Phi/2)\big)}\right)\,;
\label{eq:n_soliton} \\
u_\mathrm{s} (z;V) &=&
  V \left(1-\frac{n_0}{n_\mathrm{s}  (z;V)}\right)\,.
\label{eq:u_soliton}
\end{eqnarray}
The velocity $V$ is related to the the phase drop across the soliton
$\Phi=\phi (-\infty) -\phi(+\infty)$ as $V /c= \cos (\Phi/2)$.  One  also
finds for the momentum and energy of the soliton
\begin{eqnarray}
  \label{eq:mom_en}
  P(V) =  n_0 (\Phi-\sin \Phi)\,;\quad\quad
  E(V) = E_0 \sin^3 (\Phi/2) \,,
\end{eqnarray}
where $E_0(n_0)=4c n_0/3 $
can be regarded as the soliton chemical potential.
Expressions~(\ref{eq:mom_en})
lead to the equation of motion for a soliton moving with a constant velocity
$V=\partial_V E/\partial_V P= \partial E/\partial P$.  In what follows we
shall focus on dark, $\Phi\approx \pi$, small velocity
solitons $V\ll c$, which have $P(V)\approx \pi n_0 +M^* V$ and $E(V) \approx
E_0 + M^* V^2/2$ \cite{footvc}.

We now substitute the
soliton solution, $n_\mathrm{s} (x-X;V)$, and $u_\mathrm{s} (x-X;V)$,
Eqs.~(\ref{eq:n_soliton}),~(\ref{eq:u_soliton}), into Eq.~(\ref{eq:action}).
Here $X(t)$ is an instantaneous soliton coordinate related to the velocity $V$
by the equation of motion $\dot X =V$. As a result one  finds an
effective DS Lagrangian
\begin{eqnarray}
  \label{eq:lagrangian}
  L_s = P (\dot X)  \dot X - E(\dot X) = \pi n_0 \dot X +\frac{M^* \dot
    X^2}{2} -E_0.
\end{eqnarray}
It  provides  Feynman path integral description of the soliton
as a {\em quantum} particle with the coordinate $X(t)$ moving in
constant background with density $n(x,t)=n_0$ and $u(x,t)=0$.
Local fluctuations of density $n_0\to n_0+\rho(X,t)$ and background velocity
$u(X,t)$ interact with the soliton and modify its dynamics.
In the long-wavelength
limit  the effect of phonons can be studied by considering small changes to the
uniform  background. The change in the background density
is accounted for by  modifying the chemical potential of the soliton, the last
term in Eq.~(\ref{eq:lagrangian}), by  expanding it as
$E_0 (n_0 +\rho) \simeq E_0 + N_0(m c^2/n_0) \rho + \partial_{n_0}
(N_0 mc^2/n_0)  \rho^2/2$. Here  $N_0 =\partial E_0/\partial \mu$ is
the number of particles expelled from the dark soliton. To obtain the coupling to the velocity field, we note that
in the laboratory frame, where the liquid moves with the uniform velocity $u$,
the fields  Eqs.~(\ref{eq:n_soliton}),~(\ref{eq:u_soliton}) transform
as $n_\mathrm{s}(z,V)\to n_\mathrm{s} (z,V-u)$, $u_\mathrm{s} (z, V)\to
u+u_\mathrm{s} (z, V-u) $.  This transformation modifies  momentum
$P(V) \to P(V-u)$ and energy  $E(V) \to E(V-u) + u P(V-u)$
of the soliton due to the uniform background flow $u$ \cite{foot2}.
Together with the density corrections it leads to the
Lagrangian
\begin{eqnarray}
  \label{eq:s_int}
  L_\mathrm{s} &+& L_\mathrm{s-ph} = L_\mathrm{s} (\dot X-u, n_0+\rho)
  =
  - \pi(n_0+\rho)(\dot X -u)   \nonumber\\
  &+&  \frac{M^* \dot X^2}{2} -
  M^* \dot X u +
  \frac{M^* u^2}{2}  - 2c \rho -
  \frac{c\rho^2}{2n_0},
\end{eqnarray}
where $\rho= \rho(X,t)$ and $u= u(X,t)$.

Instead of tackling it directly, it is convenient to
perform gauge transformation of
the phonon density and velocity fields to get rid of terms linear in these
variables. This is achieved by the following substitution $\rho(x,t) \to
\rho(x,t) - N_0 \delta(x-X(t))$ along with $ u (x,t) \to u(x,t) - (\pi/m)
\delta (x-X(t))$.  One should also redefine the soliton coordinate \cite{foot3}
 $\dot X -\left(1+mN_0/M^*\right)
u(X,t) \to  \dot X$ to account for the phonon drag. Upon this
change of variables in $L_{\mathrm{ph}}+L_\mathrm{s}+L_{\mathrm{s-ph}}$
the soliton-phonon interaction Lagrangian acquires the form
\begin{eqnarray}
  \label{eq:s_int_quad}
  L_\mathrm{s-ph}  =
  - \frac{\Gamma_\rho}{2}\, \rho^2(X,t) - \frac{\Gamma_u}{2}\, u^2(X,t)\, ,
\end{eqnarray}
where the soliton -- two-phonon interaction vertices are given by
 \begin{eqnarray}
  \label{eq:gamma_rho}
  \Gamma_\rho
  = \frac{\partial \mu}{\partial n_0}\, \frac{\partial N_0}{\partial n_0}\,;
\quad\quad  \Gamma_u =  m N_0\left(1+\frac{mN_0}{M^*}\right) .
\end{eqnarray}
Notice that to derive Eqs.~(\ref{eq:s_int_quad}) and (\ref{eq:gamma_rho}) we
crucially used the phonon non-linearity $m \rho u^2/2$ in the Lagrangian
(\ref{eq:action_luttinger}).  Since we have succeeded to transform the
interaction Lagrangian to the form which does not contain terms linear in the
phonon fields, we can disregard now the phonon non-linearity and treat them as
the Luttinger liquid \cite{HaldanePRL81}, described by the Gaussian part of
Eq.~(\ref{eq:action_luttinger}) ($\rho u^2$ term does not contribute to the
leading temperature dependence).

We have arrived thus at the problem of the ''quantum impurity'' with the mass
$M^*$, interacting with the Luttinger liquid through the two-phonon vertices
(\ref{eq:s_int_quad}). This problem was considered in
Refs.~[\onlinecite{castro96}] in close analogy to three dimensional
dynamics of impurities in liquid
$^4$He \cite{LandauKhalatnikov1949ViscosityI}.
The leading $\sim T^4$ contribution  to the viscous force acting on the
``impurity'' corresponds to the process
shown in Fig.~\ref{fig:dispersion}a, where the ''impurity'' absorbs one
thermal phonon, while emitting another long
wavelength phonon to satisfy momentum and energy conservation.
It results in equation of motion  $M^* \dot V = F(V,0)$,
where the friction force exerted by the liquid on moving impurity is
\begin{eqnarray}
  \label{eq:momentum_dot}
F(V,X)\!  =\!  -\frac{1}{4} \left(\Gamma_\rho - \Gamma_u\,\frac{c^2}{n_0^2}
\right)^2\!\!\!
\sum\limits_{|q|\lesssim mc}\!\! e^{iqX} q\,
\Pi(q,qV).
\end{eqnarray}
Here $\Pi(q,\omega)$ is the imaginary part of the Fourier transform of
$\theta(t)\langle[\rho^2(x,t),\rho^2(0,0)]\rangle$ response function of
the phonon gas. For  small velocity, $V\ll c$,  one finds
\begin{equation}
  \label{eq:Pi1} \Pi (q,qV) =
\!\frac{n_0^2}{8m^2c^3 T}\,
    \frac{q^3}{\sinh^{2}(cq/4T)}\, V\,  .
\end{equation}
The momentum sum in Eq.~(\ref{eq:momentum_dot}) is limited to $|q|\lesssim
mc$.  Indeed, phonons with the wavelengths shorter than DS size
$(mc)^{-1}$ practically do not interact  with the latter.

Using the relations   $M^* = -2 m N_0 =- 4 n_0/c$ and $mc^2/n_0 = g$
  we find
$\Gamma_\rho = c/n_0$ and $ \Gamma_u = n_0/c$.
In view of Eq.~(\ref{eq:momentum_dot})
$\dot V=0$, meaning that the soliton motion is unaffected by the interactions
with the phonons.  This is a consequence of the exact integrability of the
Lagrangian in Eq.~(\ref{eq:action}), which protects the soliton from the
dissipation. This is also expected from the fact that the soliton
configuration is
an example of a reflectionless potential, playing an important role in the
classical theory of integrable models \cite{FaddeevTakhtajanBook}, so it does
not scatter phonons.  Remarkably, in our approach
this fact manifests itself through a subtle
destructive interference of the phonons excited by the density and current
vertices.

When a small cubic in density term $-\alpha(n-n_0)^3/6$ is added to the
Lagrangian (\ref{eq:action}), it breaks the exact integrability of the
problem.  Below we calculate the  corrections to the
Lagrangian (\ref{eq:s_int_quad}) and demonstrate the lack
of the exact cancelation of the prefactor in Eq.~(\ref{eq:momentum_dot}),
leading to the dissipation and eventual evaporation of the soliton
by emission of phonons. To find corrections linear in $\alpha$
to the soliton   Lagrangian (\ref{eq:lagrangian})
it is sufficient to  substitute  the bare soliton configuration,
Eqs. (\ref{eq:n_soliton})~and~(\ref{eq:u_soliton}),
into the cubic in density term,
\begin{equation}
  \label{eq:delta_s}
  \delta L_\mathrm{s}\! =\! -\frac{\alpha}{6} \int\!dx
  \, [n_s (x;V) -n_0]^3
=\! \frac{8}{45}\frac{\alpha n^3_0}{mc}
    \left(1-\frac{V^2}{c^2}\right)^\frac{5}{2}\!.
\end{equation}
Expanding in $V/c\ll 1$, and comparing with Eq.~(\ref{eq:lagrangian}), one
finds $\delta E_0 =-8\alpha n_0^3/(45mc)$ and $\delta M^*=
-8\alpha n_0^3/(9mc^3)$. To calculate corrections
to the number of particles expelled from
the soliton $N_0 = (n_0/mc^2) \partial E_0/\partial n_0$ and its derivative
which enters Eq.~(\ref{eq:gamma_rho}) it is important to
take into account the renormalization of the  sound velocity
$mc^2/n_0 = \partial\mu/\partial n_0= g +\alpha n_0$.
We obtain the modified vertices
\begin{equation}
  \label{eq:gamma_rho_alpha}
  \Gamma_\rho\! =\!
  \frac{c}{n_0}\left(1-\frac{2}{3}\frac{\alpha
      n_0^2}{mc^2} \right)\,; \quad
   \Gamma_u \!=\!
  \frac{n_0}{c}\left(1+\frac{2}{9}\frac{\alpha
      n_0^2}{mc^2}\right)\,,
\end{equation}
resulting in the non-zero soliton -- two-phonon coupling in
Eq. (\ref{eq:momentum_dot}), $\Gamma_\rho - \Gamma_uc^2/n_0^2=-
8\alpha n_0/(9mc)$.

To obtain Eqs.~(\ref{eq:vdot}), (\ref{eq:visc_low}) we notice that at small
temperature $T\ll mc^2$ the momentum sum in Eq.~(\ref{eq:momentum_dot}) may be
extended to infinity.
Substituting then Eqs.~(\ref{eq:Pi1}) and
(\ref{eq:gamma_rho_alpha}) into Eq.~(\ref{eq:momentum_dot}), one finds
$F(V,X)=-\kappa(T)V f(X/L_T)$, where
\begin{eqnarray}
  \label{eq:f}
  f\left(\frac{y}{2\pi}\right)  = \frac{30}{\pi^{4}} \int\limits d k \,
\frac{k^4 \cos(2ky/\pi)}{\sinh^{2} k}
  =\frac{15}{\sinh^5y}\nonumber\\
   \times\Big[ \sinh y\big( 3 +2\sinh^2y\big)
- y \cosh y
    \left(2+\cosh^2y\right) \Big].
\end{eqnarray}
The $T^4$ dependence of the friction force  due to ``cold'' phonons with
$q<T/c\ll mc$ \cite{castro96} is to be contrasted to the
regime of larger temperatures $T\gg \mu\sim mc^2$ (still $T\ll E_0 $) where
all the phonons with $q\lesssim mc$ contribute to the dissipation,
their number being simply proportional to $T$. One finds thus $\kappa\sim
(\alpha n_0^2/mc^2)^2 m T$ in agreement with the results of
Ref.~[\onlinecite{Muryshev2002Dynamics}].  The numerical coefficient here
depends on details of the momentum cutoff at $q\sim mc$ and is beyond our
phenomenological long wavelength approach.
To derive the equations of motion (\ref{eq:soliton-gas}) for the gas of DS we
describe interaction of each soliton with the phonons by the Lagrangian
(\ref{eq:s_int_quad}).  Integrating out the Gaussian phonons and taking the
semiclassical limit, we arrive at Eq.~(\ref{eq:soliton-gas}) with the force
given by Eq.~(\ref{eq:f}).

In conclusion,  soliton scattering on background fluctuations,
leads to its decay and a finite lifetime
(unless the model is integrable). We
have shown that in the quantum regime such a lifetime is significantly
longer, than expected from classical considerations and solitons acquire
long-range mutual interactions. Our approach may prove useful in other
areas studying solitons propagation in a dynamic media, such as, e.g.,
non-linear optics.

The authors are grateful to M.~Pustilnik for discussions which have
initiated this project and to K.~Bongs, M.~K\"ohl and K.~Matveev
for fruitful discussions.
D.M.G.  acknowledges the hospitality of
Prof. Dr. K. Sengstock's group in Hamburg and
support by EPSRC Advanced Fellowship EP/D072514/1.
A.K.  was supported by NSF grant DMR-0804266 and EPSRC grant GR/T23725/01.


\begin{thebibliography}{14}
\expandafter\ifx\csname natexlab\endcsname\relax\def\natexlab#1{#1}\fi
\expandafter\ifx\csname bibnamefont\endcsname\relax
  \def\bibnamefont#1{#1}\fi
\expandafter\ifx\csname bibfnamefont\endcsname\relax
  \def\bibfnamefont#1{#1}\fi
\expandafter\ifx\csname citenamefont\endcsname\relax
  \def\citenamefont#1{#1}\fi
\expandafter\ifx\csname url\endcsname\relax
  \def\url#1{\texttt{#1}}\fi
\expandafter\ifx\csname urlprefix\endcsname\relax\def\urlprefix{URL }\fi
\providecommand{\bibinfo}[2]{#2}
\providecommand{\eprint}[2][]{\url{#2}}

\bibitem[{\citenamefont{Konotop and Pitaevskii}(2004)}]{Konotop2004Landau}
  W.P.~Reinhardt and C.W.~Clark, J. Phys. B \textbf{30}, L785 (1997);
Th.~Busch and J.R.~Anglin, Phys. Rev. Lett. \textbf{84}, 2298 (1999);
  D.J.~Frantzeskakis \textit{et al.}, Phys. Rev. A \textbf{66}, 053608 (2002);
\bibinfo{author}{\bibfnamefont{V.~V.} \bibnamefont{Konotop}} \bibnamefont{and}
  \bibinfo{author}{\bibfnamefont{L.}~\bibnamefont{Pitaevskii}},
  \bibinfo{journal}{Phys. Rev. Lett.} \textbf{\bibinfo{volume}{93}}
  (\bibinfo{year}{2004});
  N.~Bilas and N.~Pavloff, Phys. Rev. A \textbf{72}, 033618 (2005);

\bibitem[{\citenamefont{Muryshev et~al.}(2002)\citenamefont{Muryshev,
  Shlyapnikov, Ertmer, Sengstock, and Lewenstein}}]{Muryshev2002Dynamics}
\bibinfo{author}{\bibfnamefont{A.}~\bibnamefont{Muryshev}} \textit{et~al.},
  \bibinfo{journal}{Phys. Rev. Lett.} \textbf{\bibinfo{volume}{89}},
  \bibinfo{pages}{110401} (\bibinfo{year}{2002}).

\bibitem{Jackson2007} B.~Jackson, N.P.~Proukakis, and C.F.~Barenghi,
  Phys. Rev. A \textbf{75}, 051601(R) (2007).


\bibitem[{\citenamefont{Burger et~al.}(1999)\citenamefont{Burger, Bongs,
  Dettmer, Ertmer, Sengstock, Sanpera, Shlyapnikov, and
  Lewenstein}}]{Burger1999Dark}
\bibinfo{author}{\bibfnamefont{S.}~\bibnamefont{Burger}} \textit{et~al.},
  \bibinfo{journal}{Phys. Rev. Lett.} \textbf{\bibinfo{volume}{83}},
  \bibinfo{pages}{5198} (\bibinfo{year}{1999}); J.~Denshclag \textit{et al.},
  Science \textbf{287}, 97 (2000); B.~Anderson \textit{et al.},
  Phys. Rev. Lett. \textbf{86}, 2926 (2001).

\bibitem[{\citenamefont{Becker et~al.}(2008)\citenamefont{Becker, Stellmer,
  Soltan-Panahi, D\"orcher, Baumert, Richter, Kronj\"ager, Bongs, and
  Sengstock}}]{Becker08}
\bibinfo{author}{\bibfnamefont{C.}~\bibnamefont{Becker}} \textit{et~al.},
  \bibinfo{journal}{Nature Physics} \textbf{\bibinfo{volume}{4}},
  \bibinfo{pages}{496} (\bibinfo{year}{2008}).


\bibitem{Dimensional_crossover} A.D.~Jackson, G.M.~Kavoulakis and
  C.J.~Pethick, Phys. Rev. A \textbf{58} 2417 (1998); A.E.~Muryshev, H.B. van
  Linden van den Heuvell, and G.V.~Shlyapnikov, Phys. Rev. A \textbf{60},
  R2665 (1999);
  D.L.~Feder
  \textit{et al.}, Phys. Rev. A \textbf{62}, 053606 (2000); J.~Brand and
  W.P.~Reinhardt, Phys. Rev. A \textbf{65}, 043612 (2002);
  G.~Theocharis
  \textit{et al.}, Phys. Rev. A \textbf{76}, 045601 (2007).




\bibitem[{\citenamefont{Lieb and Liniger}(1963)}]{Lieb_Liniger_1963}
\bibinfo{author}{\bibfnamefont{E.~H.} \bibnamefont{Lieb}} \bibnamefont{and}
  \bibinfo{author}{\bibfnamefont{W.}~\bibnamefont{Liniger}},
  \bibinfo{journal}{Phys. Rev.} \textbf{\bibinfo{volume}{130}},
  \bibinfo{pages}{1605} (\bibinfo{year}{1963}).

\bibitem[{\citenamefont{Kulish et~al.}(1976)\citenamefont{Kulish, Manakov, and
  Faddeev}}]{KulishManakovFaddeev76}
\bibinfo{author}{\bibfnamefont{P.}~\bibnamefont{Kulish}},
  \bibinfo{author}{\bibfnamefont{S.}~\bibnamefont{Manakov}}, \bibnamefont{and}
  \bibinfo{author}{\bibfnamefont{L.}~\bibnamefont{Faddeev}},
  \bibinfo{journal}{Theor. Math. Phys.} \textbf{\bibinfo{volume}{28}},
  \bibinfo{pages}{615} (\bibinfo{year}{1976}).

\bibitem[{\citenamefont{Lieb}(1963)}]{Lieb_1963}
\bibinfo{author}{\bibfnamefont{E.~H.} \bibnamefont{Lieb}},
  \bibinfo{journal}{Phys. Rev.} \textbf{\bibinfo{volume}{130}},
  \bibinfo{pages}{1616} (\bibinfo{year}{1963}).

\bibitem[{\citenamefont{Faddeev and Takhtajan}(1987)}]{FaddeevTakhtajanBook}
\bibinfo{author}{\bibfnamefont{L.~D.} \bibnamefont{Faddeev}} \bibnamefont{and}
  \bibinfo{author}{\bibfnamefont{L.~A.} \bibnamefont{Takhtajan}},
  \emph{\bibinfo{title}{Hamiltonian Methods in the Theory of Solitons}}
  (\bibinfo{publisher}{Springer}, \bibinfo{address}{Berlin, Heidelberg},
  \bibinfo{year}{1987}).

\bibitem{Dziarmaga} J.~Dziarmaga, Phys. Rev. A \textbf{70} 063616 (2004).

\bibitem{Moelmer} A.~Negretti, C.~ Henkel, and K. M{\o}lmer, Phys. Rev. A \textbf{78} 023630
  (2008).


\bibitem[{\citenamefont{Mazets et~al.}(2008)\citenamefont{Mazets, Schumm, and
  Schmiedmayer}}]{Mazets2008Breakdown}
\bibinfo{author}{\bibfnamefont{I.~E.} \bibnamefont{Mazets}},
  \bibinfo{author}{\bibfnamefont{T.}~\bibnamefont{Schumm}}, \bibnamefont{and}
  \bibinfo{author}{\bibfnamefont{J.}~\bibnamefont{Schmiedmayer}},
  \bibinfo{journal}{Phys. Rev. Lett.} \textbf{\bibinfo{volume}{100}}
  \bibinfo{pages}{210403} (\bibinfo{year}{2008}).



  \bibitem{foot1} According to  fluctuation-dissipation theorem,
the friction force is accompanied by  stochastic force $\xi(t)$
with the correlator
$\langle \xi(t)\xi(t')\rangle = 2\kappa T\delta(t-t')$.  The complete
semiclassical description of the soliton dynamics  is given by the
distribution function, satisfying the Fokker-Planck
equation. Here we restrict ourselves with the dynamics of the maximum of this
distribution only.

\bibitem{footvc} Eq.~(\ref{eq:tau}), based on
Eqs.~(\ref{eq:vdot}),~(\ref{eq:visc_low})
is valid  in the experimentally most relevant regime $V\ll c$.
A more general case,
which includes the KdV regime $V\lesssim c$ is treated elsewhere
\cite{GangardtKamenevInPreparation}.


\bibitem[{\citenamefont{Pitaevskii and
  Stringari}(2003)}]{PitaevskiiStringariBook}
\bibinfo{author}{\bibfnamefont{L.~P.} \bibnamefont{Pitaevskii}}
  \bibnamefont{and}
  \bibinfo{author}{\bibfnamefont{S.}~\bibnamefont{Stringari}},
  \emph{\bibinfo{title}{Bose-Einstein Condensation}}
  (\bibinfo{publisher}{Clarendon Press}, \bibinfo{address}{Oxford},
  \bibinfo{year}{2003}).


\bibitem{foot2}  This transformation shows that DS behaves  as a particle with
zero bare mass $M=0$, but finite effective mass $M^*$.

\bibitem{foot3} The corresponding transformation of the arguments of $\rho(X,t)$ and
$u(X,t)$ leads to higher order interactions,  not contributing to
the leading small temperature result.


\bibitem[{\citenamefont{Haldane}(1981)}]{HaldanePRL81}
\bibinfo{author}{\bibfnamefont{F.~M.~D.} \bibnamefont{Haldane}},
  \bibinfo{journal}{Phys. Rev. Lett.} \textbf{\bibinfo{volume}{47}},
  \bibinfo{pages}{1840} (\bibinfo{year}{1981}).

\bibitem[{\citenamefont{Castro~Neto and P.A.}(1996)}]{castro96}
\bibinfo{author}{\bibfnamefont{A.~H.} \bibnamefont{Castro~Neto}}
  \bibnamefont{and} \bibinfo{author}{ \bibnamefont{M.P.A. Fisher}},
  \bibinfo{journal}{Phys. Rev. B} \textbf{\bibinfo{volume}{53}},
  \bibinfo{pages}{9713} (\bibinfo{year}{1996});
\bibinfo{author}{\bibfnamefont{D.~M.} \bibnamefont{Gangardt}} \bibnamefont{and}
  \bibinfo{author}{\bibfnamefont{A.}~\bibnamefont{Kamenev}},
  \bibinfo{journal}{Phys. Rev. Lett.} \textbf{\bibinfo{volume}{102}},
  \bibinfo{pages}{070402} (\bibinfo{year}{2009}).

\bibitem{LandauKhalatnikov1949ViscosityI}
\bibinfo{author}{\bibfnamefont{L.~D.} \bibnamefont{Landau}} \bibnamefont{and}
  \bibinfo{author}{\bibfnamefont{I.~M.} \bibnamefont{Khalatnikov}},
  \bibinfo{journal}{Zh. Eksp. Teor. Fiz.} \textbf{\bibinfo{volume}{19}},
  \bibinfo{pages}{637} (\bibinfo{year}{1949}{\natexlab{a}}).


\bibitem{GangardtKamenevInPreparation} D.M.~Gangardt and A.~Kamenev, in
  preparation.

\end{thebibliography}

\end{document}